\documentclass[aps, prl, twocolumn, showpacs, showkeys, superscriptaddress, nofootinbib]{revtex4} 
\usepackage{graphics} 
\usepackage{amssymb} 
\usepackage{amsmath} 
\usepackage{amsfonts} 
\usepackage{mathrsfs} 
\begin{document} 
\title{Events and observables in generally invariant spacetime theories} 
\author{Hans Westman\footnote{\tt hwestman@perimeterinstitute.ca}} 
\affiliation{Perimeter Institute for Theoretical Physics, 31 Caroline Street N., Waterloo, Ontario, N2L 2Y5 Canada} 
\author{Sebastiano Sonego\footnote{\tt sebastiano.sonego@uniud.it}} 
\affiliation{Universit\`a di Udine, Via delle Scienze 208, 33100 Udine, Italy} 
\date{June 30, 2008; \LaTeX-ed \today} 
\bigskip 
\begin{abstract} 
We address the problem of observables in generally invariant spacetime theories such as Einstein's general relativity.  Using the refined notion of an event as a ``point-coincidence'' between scalar fields that completely characterise a spacetime model, we propose a generalisation of the relational local observables that does not require the existence of four everywhere invertible scalar fields.  The collection of all point-coincidences forms in generic situations a four-dimensional manifold, that is naturally identified with the physical spacetime. 
\end{abstract} 
\maketitle 
\def\g{\mbox{\sl g}} 
\def\eg{{\em e.g.\/}} 
\def\ie{{\em i.e.\/}} 

\noindent\underline{\em Introduction\/}  
In spacetime theories~\cite{friedman}, a model of the physical world is usually a pair $({\cal M},T)$, where $\cal M$ is a four-dimensional manifold with suitable topological and differentiable properties, and $T$ represents a collection of tensor fields on $\cal M$.  Points of $\cal M$ are commonly interpreted as events, which are labeled by a suitable set of coordinates $x^\mu$ in a chart.  (Greek indices $\mu,\nu,\ldots$ run from 1 to 4.)  The value of some physical quantity, represented by a scalar field $\phi$ on $\cal M$, at a point $p\in {\cal M}$ is then given by $\phi(p)=\bar{\phi}(x^1,\ldots,x^4)$, where $\bar{\phi}$ is the representation of $\phi$ in the chart one is working with.  Na\"{\i}vely, the function $\bar{\phi}(x^1,\ldots,x^4)$ should correspond to an observable, giving the value of the field $\phi$ at the event labeled by the coordinates $(x^1,\ldots,x^4)$, for all events in a given spatiotemporal region.  However, such an interpretation relies on the assumption that, in addition to the value of the field, {\em also\/} those of the coordinates $x^1,\ldots,x^4$ are operationally well-defined quantities. This is the case in theories, like Newtonian mechanics and special relativity, that (implicitly) postulate the existence of physical systems whose behaviour is independent of, and does not affect, the phenomena one wishes to observe.  Such systems can be used to construct physical reference frames on which one can read directly the values of four variables $x^1,\ldots,x^4$, that so acquire operational significance. However, it turns out that in theories like general relativity, whose equations are invariant under arbitrary coordinate transformations, one cannot give such an operational interpretation to the coordinates, which must then be regarded as mere mathematical parameters, devoid of any physical meaning.  In these theories, it is therefore neither obvious how one should characterise observable quantities, nor how spacetime --- the set of all events --- is described.  In this letter, we present the line of arguments that leads to these problems, and outline a possible resolution.  Our construction of the set of events for an arbitrary model, and of the corresponding observables, is not only potentially useful in the programme of the quantisation of gravity but, intriguingly, it also clarifies the ontology of classical spacetime, and sheds light on several rather cryptical statements made by Einstein.  We deliberately avoid making extensive connections to the extant literature, both interpretive and technical, in order not to distract the attention of the reader from the main logical flow.  A more thorough account is given in a companion paper~\cite{companion}, where we also deal with many side issues that one encounters in this type of investigations.

\noindent\underline{\em General invariance\/}  
Consider a set of field equations for the collection $T$ of tensor fields in a spacetime theory.  When these equations are written down explicitly for the components of $T$ in a chart with coordinates $x^\mu$, they are a set of partial differential equations in the independent variables $x^\mu$.  Consider now the coordinate transformation $x^\mu\to x'^\mu$, defined by $x^\mu=f^\mu(x')$, with the $f^\mu$ differentiable functions. The components of $T$ in the new chart obey, of course, new partial differential equations in the independent variables $x'^\mu$.  If these equations look exactly the same (except for trivial relabeling) as those satisfied by the components of $T$ in the old chart with respect to the variables $x^\mu$, the transformation $x^\mu\to x'^\mu$ corresponds to a symmetry of the theory.  If the set of equations is such that {\em every\/} $f^\mu$ corresponds to a symmetry, we say that the theory is {\em generally invariant\/}.

This is a nontrivial property of a system of differential equations, and should not be confused with the mathematical possibility of formulating a theory using tensors --- a property with little physical relevance~\cite{kretschmann}, that we shall denote {\em general covariance\/}.\footnote{Failing to appreciate the difference between invariance (a symmetry property of a set of equations) and covariance (a property of the formal apparatus used in a physical theory) has produced a huge literature.  See reference~\cite{norton} for a thorough review, and reference~\cite{giulini} for a clear and mathematically clean presentation of the concepts.} Of course, coordinates may not always cover the whole spatiotemporal region of interest.  However, this poses no problem for the previous definition of invariance:  The field equations can be restricted to a topologically trivial open set of $\cal M$, and invariance can be checked for all such open regions.

Einstein's theory of gravity is generally invariant.  For example, in an empty region of spacetime Einstein's equations for the metric components $\g_{\mu\nu}(x)$ and for 
\begin{equation} 
\g'_{\mu\nu}(x')=\frac{\partial f^\rho(x')}{\partial x'^\mu}\,
\frac{\partial f^\sigma(x')}{\partial x'^\nu}\,
\g_{\rho\sigma}(f(x')) 
\label{gg'} 
\end{equation} 
are exactly the same, apart from the choice of symbols ($x$ and $\g_{\mu\nu}$ {\em versus\/} $x'$ and $\g'_{\mu\nu}$).  This property holds also in the presence of matter, provided that one adds suitable field equations for the non-gravitational fields as well.  For the sake of definiteness, hereafter we shall focus on Einstein's theory; however, all the discussion can be easily adapted to an arbitrary generally invariant theory.

One can use general invariance to generate solutions of Einstein's equations.  Let $\g_{\mu\nu}(x)$ be a solution in the coordinates $x^\mu$.  Then $\g'_{\mu\nu}(x')$ is a solution in the coordinates $x'^\mu$, for all choices of $f$.  On replacing $x'$ by $x$, and using general invariance, we have that the new functions 
\begin{equation} 
\g'_{\mu\nu}(x)=\frac{\partial f^\rho(x)}{\partial x^\mu}\,
\frac{\partial f^\sigma(x)}{\partial x^\nu}\,
\g_{\rho\sigma}(f(x)) 
\label{new-g} 
\end{equation} 
solve Einstein's equations in the coordinates $x^\mu$.

\noindent\underline{\em Coordinates are not readings\/}  
Let $\g_{\mu\nu}(x)$ be a solution of Einstein's equations in some coordinates $x^\mu$.  Within a suitable open region of the coordinate domain, $\g_{\mu\nu}(x)$ can be regarded as the solution to an initial value problem formulated on a three-dimensional hypersurface $\cal S$.  We can now use the general invariance of Einstein's equations to generate, starting from $\g_{\mu\nu}(x)$, a different solution $\g'_{\mu\nu}(x)$ that satisfies the same initial value problem.  For this purpose, it is sufficient to choose the functions $f^\mu$ in Eq.~(\ref{new-g}) such that they coincide with the identity in a neighbourhood of $\cal S$, while they differ from it elsewhere.

This simple remark has far-reaching consequences~\cite{friedman, earman}.  Let us adopt an interpretation of coordinates in which they represent physical readings (as it is common in Newtonian mechanics and in special relativity).  Then, one expects any measurable quantity (for example, a curvature scalar), to be expressed by a unique well-defined scalar function of the $x^\mu$, say $\phi(x)$.\footnote{With some abuse of notation, hereafter we shall denote by the same symbol both a function defined on $\cal M$ and its coordinate representation in a chart, since there is no possibility of confusion.}  Indeed, in any actual experiment only one correspondence between the values of that quantity and those of the (by assumption) operationally well-defined readings $x^\mu$ will be found.  However, as stated above, $\g_{\mu\nu}(x)$ and $\g'_{\mu\nu}(x)$ are both solutions to the same initial value problem, and in these two mathematically distinct solutions the functional dependence of any scalar on the coordinates is $\phi(x)$ and $\phi'(x)=\phi(f(x))$, respectively.  Since $\phi(x)\neq\phi(f(x))$ in general and so $\phi(x)\neq\phi'(x)$, it follows that, because of the general invariance of Einstein's equations, general relativity does not predict a unique value of $\phi$ for given values of the operationally well-defined quantities $x^\mu$, and the theory is thus unable to make unique empirical predictions.  This is an untenable conclusion, so the operational interpretation of the coordinates must be dropped, by a {\em reductio ad absurdum\/}.

Summarising, unless one is ready to accept the lack of unique empirical predictions, one cannot assume that the coordinates $x^\mu$ have any operational meaning --- that is, that they correspond to readings of some sort, identifying a well-defined position in space and time.  They are just {\em mathematical parameters\/}.  Note that this should not be interpreted in the trivial sense that charts on a manifold are arbitrary, because, given a chart, there is a one-to-one correspondence between a manifold point and the coordinates.  What we are saying is actually that the {\em points of\/} $\cal M$ lack operational significance, \ie, they do not represent operationally well-defined events.\footnote{This important conceptual point is seldom made in the literature, and particularly in textbooks, where $\cal M$ is often presented as the set of events (defined by physical occurrences) and the mathematical coordinates are identified with physical readings of some sort.}  The manifold $\cal M$ must be thought of as a purely abstract space, whose points possess no physical quality that could allow one to identify them.

\noindent\underline{\em Observables\/}  
An immediate consequence of the previous conclusion is that in general relativity one cannot experimentally establish a functional relation between the values of the parameters $x^\mu$ and those of physical quantities, because there is no way of ``reading'' the values of the $x^\mu$. Hence, statements referring to the value of a given field at some point of $\cal M$ are, if taken literally, physically empty, and observable quantities cannot, in general, be represented by functions on $\cal M$.  In order to extract observables from a given spacetime model $({\cal M},T)$ we must therefore, in one way or another, eliminate the coordinate dependence.  We now outline a particular way of doing that, which leads to the so-called relational local observables~\cite{bergmann}.

In every given spacetime model $({\cal M},T)$, although the correspondence between manifold parameters (the $x^\mu$) and values of physical fields is not observable, the correspondence between values of physical fields and values of other physical fields {\em is\/} physically meaningful.  Indeed, this correspondence contains {\em everything\/} one can measure.  To define observable quantities, we then can construct, out of all the fields $T$ in the model, four scalar {\em coordinate fields\/}\footnote{Indices $\alpha,\beta,\ldots$ run from 1 to 4, but label scalar quantities and should not be confused with the tensor indices $\mu,\nu,\ldots$.} $q^\alpha$, and express any other quantity in terms of these. We assume that these four scalar fields are invertible, \ie, that in some open set ${\cal U}\subseteq {\cal M}$ one has $\det\left(\partial q^\alpha/\partial x^\nu\right) \neq 0$.  We can then define a one-to-one map $q:=(q^1,\ldots,q^4)$ of $\cal U$ onto a subset ${\cal Q}\subseteq\mathbb{R}^4$ with non-zero measure, so $q:{\cal U}\to {\cal Q}$ and $q^{-1}:{\cal Q}\to {\cal U}$ are both well-defined.

Consider any scalar function $\phi:{\cal M}\to \mathbb{R}$.  By composing $\phi$ and $q^{-1}$ we obtain the function $\tilde{\phi} =\phi\circ q^{-1}:{\cal Q}\to \mathbb{R}$, or using coordinates $\tilde{\phi}(q)=\phi(x(q))$, where $x(q)$ denotes the values of the coordinates corresponding to the values $q$ of the physical coordinate fields.  This no longer contains associations between the measurable field values and the unobservable points of ${\cal M}$ (or coordinates $x^\mu$), but only between measurable field values and other measurable quantities --- the $q^\alpha$.  Hence, contrary to what happened for the field $\phi$ on $\cal M$ (or its coordinate representation), the function $\tilde{\phi}(q)$ is observable and is the same in all models $({\cal M'},T')$ related to $({\cal M},T)$ by a diffeomorphism.  Objects constructed in this way are also Dirac observables within the canonical framework~\cite{Dittrich06}.

In order to construct observables from tensor quantities, one first defines the four one-forms ${e^\alpha}_\mu:=\partial q^\alpha/\partial x^\mu$.  Because of invertibility, $\det({e^\alpha}_\mu)\neq 0$ and one can also introduce four vectors ${f_\alpha}^\mu$ such that ${e^\alpha}_\mu {f_\beta}^\mu=\delta^\alpha_\beta$.   Now we can use these tetrads to construct scalars out of tensorial objects.  For example: 
\begin{equation}
\mbox{$R_{\alpha\beta\gamma}$}^\delta(x)={f_\alpha}^\mu(x)
{f_\beta}^\nu(x) {f_\gamma}^\rho(x) {e^\delta}_\sigma(x)
{R_{\mu\nu\rho}}^\sigma(x)\;. 
\label{threescalars} 
\end{equation} 
Such scalars depend on the coordinates, so they are not yet observable quantities.  However, observables can now easily be constructed as already discussed, by making use of the relation $x=x(q)$: 
\begin{equation} 
{\widetilde{R}{_{\alpha\beta\gamma}}}^\delta(q)
=\mbox{$R_{\alpha\beta\gamma}$}^\delta(x(q))\;. 
\end{equation} 
All the local observable quantities of the theory can be generated in this way, possibly switching to other coordinate fields whenever invertibility fails. From them, one can read directly the values of physical quantities corresponding to the measured values of the coordinate fields.

An essential assumption for the viability of the above construction of local observables is that the function $q:{\cal U}\to {\cal Q}$ be invertible.  But this is of course not guaranteed by any physical law.  In fact, this hypothesis can be satisfied only locally, and only {\em once\/} one has a specific model of spacetime.  There is no way to choose {\em a priori\/} four fields $q^\alpha$ that can be used everywhere in ${\cal M}$ for a given model, and for all models.  This is, of course, not too problematic for classical general relativity, but in a quantum theory of gravity it could be that no spacetime model is specified, so the local observables now introduced are ill-defined and therefore not suitable for being turned into operators.

\noindent\underline{\em The space of point-coincidences\/}  
We now outline a way to construct local observables which does not suffer from the above problem of invertibility.  The root of the problem lies in the fact that some scalar fields (the $q^\alpha$), are selected to play a special role, so in the following we shall treat all dynamical degrees of freedom ``democratically''.  At the same time, we shall solve a puzzling foundational question that naturally arises once the ``readings interpretation'' of coordinates is rejected: If events cannot be identified with points of the manifold ${\cal M}$, how are they represented in a generally invariant spacetime theory?

Suppose that one can, from a given model $({\cal M},T)$, construct a new one $({\cal M},\Phi^1,\ldots,\Phi^N)$, where $\Phi^1,\ldots,\Phi^N$ are scalars which completely characterise the model.\footnote{Such scalars need not be fundamental fields; they could well correspond to phenomenological properties (see reference~\cite{companion} for examples).  Note that the use of scalars is mandatory: One cannot use tensor functions because they are not real-valued; nor their coordinate representations, which are not chart-independent.}  The notion of an event can be refined into the one of a ``point-coincidence''~\cite{kretschmann, einstein-1916}, defined by the concomitant values of all these scalars.  Considering the map $\mbox{\boldmath $\Phi$} := (\Phi^1,\ldots,\Phi^N) : {\cal M}\to\mathbb{R}^N$, we can define the {\em space of point-coincidences\/} ${\cal E} := \mbox{\boldmath $\Phi$}({\cal M}) \subset\mathbb{R}^N$, which we take as the formal representation of spacetime.  This is a set of ordered $N$-tuples of real numbers, which are however not all independent, because the rank of the $N\times 4$ matrix of their derivatives with respect to the $x^\mu$ cannot be greater than 4.  

In fact, if for all pairs of points of $\cal M$ at least one of the functions $\Phi^1,\ldots,\Phi^N$ takes different values, then $\cal E$ is also a manifold, with the same dimension as $\cal M$.  The other possibility arises, for example, if the model $({\cal M},T)$ possesses some symmetry (of {\em all\/} the fields in the collection $T$).  However, such cases certainly do not correspond to the spacetime of our experience and should be regarded as pathological; so this problem is far less serious than the one of invertibility mentioned at the end of the previous section.  Thus, in a generic model the values of the scalars $\Phi^1,\ldots,\Phi^N$ are constrained by conditions of the type 
\begin{equation} 
F_A(\Phi^1,...,\Phi^N)=0\;, 
\label{FA} 
\end{equation} 
where $A$ runs from 1 to $N-4$. In generic situations, these conditions define a $4$-dimensional submanifold in $\mathbb{R}^N$.

The map $\mbox{\boldmath $\Phi$}:{\cal M}\to \mathbb{R}^N$ is essentially a parametrisation of $\cal E$. Instead of using parameters/coordinates to mathematically characterise the totality of point-coincidences, one can characterise it implicitly through equations (\ref{FA}).  Thereby, the use of coordinates is completely eliminated and one is left only with structure that is empirically accessible (at least in principle).

\noindent\underline{\em Ontology of spacetime\/}  
The space of point-coincidences $\cal E$ is a four-dimensional manifold in generic situations, contains all local observable data, and its elements represent physical events (which are always characterised by concrete properties).  It is therefore natural to identify $\cal E$ with spacetime itself.  This choice is consistent with the intuitive notion of spacetime as the collection of all events, but is at variance with much of the extant literature, in which spacetime is simply identified with the manifold $\cal M$.  However, as we have seen, points of $\cal M$ are not empirically observable, in contrast with events.

On the other hand, if one identifies spacetime with the space of point-coincidences, its points (the events) are defined through observable properties of the physical and geometrical fields.  Hence, spacetime is a collection of properties of the fields, rather than a ``container'' physical objects are {\em in\/}.  This is probably the meaning of the following claims of Einstein's~\cite{einstein}: 
\begin{quote} 
{\em There is no such thing as an empty space, i.e., a space without field.  Space-time does not claim existence on its own, but only as a structural property of the field.\/} 
\end{quote} 
\begin{quote} 
{\em Physical objects are not in space, although they are spatially extended.\/} 
\end{quote} 
Indeed, without fields there is no spacetime --- a trivial statement within the view we have developed, since point-coincidences are defined only in terms of field values; but a puzzling one if one thinks, incorrectly, that fields live in a spacetime arena.  In particular, it makes no sense to think of a region of spacetime where there are no fields at all (no electromagnetic field, no scalar field, etc., and in particular no metric field).  That region would simply not exist at all.

\noindent\underline{\em Conclusions\/}  
According to the view presented in this letter, spacetime should not be thought of as a primitive, independently existing entity --- a ``container'' in which fields and the metric ``live''.  We have seen that, if the field equations are generally invariant, the manifold $\cal M$ cannot represent something empirically accessible; so, in particular, it cannot represent spacetime.  We have also seen that there is an alternative, and more natural, representation of the latter: The space of point-coincidences $\cal E$, which is the totality of physical events and contains all the observables of the theory.  However, $\cal E$ is constructed out of the fields themselves, so its existence cannot be postulated before any given field configuration is assigned.  Indeed, everything that is important in this view are the mutual relationships of the configurations of various fields --- a conception that can be regarded as corresponding to some kind of relational ontology.  The situation is very well summarised by Einstein~\cite{jammer}: 
\begin{quote} 
{\em [...] the whole of physical reality could perhaps be represented as a field whose components depend on four space-time parameters.  If the laws of this field are in general covariant, that is, are not dependent on a particular choice of co\"ordinate system, then the introduction of an independent (absolute) space is no longer necessary.  That which constitutes the spatial character of reality is then simply the four-dimensionality of the field.  There is then no ``empty'' space, that is, there is no space without a field.\/} 
\end{quote} 
(Note that Einstein uses the singular ``field'' instead of our ``fields'', probably because of his belief in a unified field theory.)

Identifying spacetime with the space $\cal E$ of point-coincidences offers a possibility for making precise some notions that one sometimes encounters in the literature about quantum gravity, such as ``fuzzy spacetime'' or ``fractal spacetime''.  Indeed, from the perspective here developed it is somewhat unnatural that the set $\cal E$ should behave as a four-dimensional smooth manifold everywhere and at every resolution.

\smallskip 


\begin{thebibliography}{99}
\bibitem{friedman} 
M.~Friedman,  
{\em Foundations of Space-Time Theories\/}  
(Princeton University Press, Princeton, 1983). 
\bibitem{companion} 
H.~Westman and S.~Sonego, 
``Coordinates, observables and symmetry in relativity'',  
Ann.\ Phys.\ (N.Y.) {\bf 324}, 1585--1611 (2009);  
e-print 0711.2651 [gr-qc]. 
\bibitem{kretschmann}
E.~Kretschmann,  
``\"Uber den physikalischen Sinn der Relativit\"atspostulate, A.\ Einstein neue und seine urspr\"ungliche Relativit\"atstheorie'',  
Ann.\ Phys.\ (Leipzig) {\bf 53}, 575--614 (1917). 
\bibitem{norton}
 J.~D.~Norton,  
``General covariance and general relativity: eight decades of dispute'',  
Rep.\ Prog.\ Phys.\ {\bf 56}, 791--858 (1993). 
\bibitem{giulini}
D.~Giulini,  
 ``Some remarks on the notions of general covariance and background  independence'',  
in {\em Approaches to Fundamental Physics:  An Assessment of Current Theoretical Ideas\/}, 
edited by E.~Seiler and I.-O.~Stamatescu, 
Lecture Notes in Physics {\bf 721}, 105--120 (Springer, Berlin, 2007); 
e-print gr-qc/0603087. 
\bibitem{earman}
J.~Earman and J.~D.~Norton,  
``What price substantivalism?  The hole story'',  
British J.\ Phil.\ Sci.\ {\bf 38}, 515--525 (1987).  
J.~Earman, 
{\em World Enough and Space-Time\/}  
(Cambridge University Press, Cambridge, 1989). 
\bibitem{bergmann} 
P.\ G.\ Bergmann,  
``Observables in general relativity'',  
Rev.\ Mod.\ Phys.\ {\bf 33}, 510--514 (1961).  
B.~S.~DeWitt, %
``The quantization of geometry'', 
in {\em Gravitation: An Introduction to Current Research\/}, 
edited by L.\ Witten  
(Wiley, New York, 1962), pp.\ 266--381.  
C.~Rovelli,  
``What is observable in classical and quantum gravity?'',  
Class.\ Quantum Grav.\  {\bf 8}, 297--316 (1991);  
{\em Quantum Gravity\/} 
(Cambridge University Press, Cambridge, 2004). 
\bibitem{Dittrich06} 
B.~Dittrich,  
``Partial and complete observables for canonical general relativity'',  
Class.\ Quantum Grav.\ {\bf 23}, 6155--6184 (2006);  
e-print gr-qc/0507106. 
\bibitem{einstein-1916} 
A.~Einstein,  
``Die Grundlagen der allgemeinen Relativit\"atstheorie'',  
Ann.\ Phys.\ (Leipzig) {\bf 49}, 769--822 (1916);  
English translation in  
{\em The Principle of Relativity\/},  
edited by A.~Sommerfeld, W.~Perrett, and G.~B.~Jeffery  
(Dover, New York, 1952), pp.~109--164.
\bibitem{einstein}
A.~Einstein,  
``Relativity and the problem of space'',  
Appendix V to {\em Relativity: The Special and General Theory\/}  
(Crown, New York, 1961), pp.\ 135--157, and preface
to the 15th edition. 
\bibitem{jammer}
A.~Einstein,  
foreword to {\em Concepts of Space\/},  
by M.\ Jammer  
(Harvard University Press, Cambridge, 1954). 
\end{thebibliography}
\end{document}